\documentclass[aps,prb,reprint,superscriptaddress,amsmath,amssymb,floatfix,showpacs]{revtex4-1}
\usepackage{mathtools}
\usepackage{amsmath}
\usepackage{graphicx}
\usepackage{esint}

\makeatletter
\usepackage{color}
\usepackage{epstopdf}
\usepackage{slashed}
\usepackage{bm}
\usepackage{braket}
\usepackage{soul}

\makeatother

\begin{document}

\title{Semiclassical ground-state phase diagram and multi-$\mathbf{Q}$
phase of a spin-orbit coupled model on triangular lattice}

\author{Changle Liu}

\affiliation{Department of Physics, Renmin University of China, Beijing 100872,
China }

\author{Xiaoqun Wang}

\affiliation{Department of Physics and Astronomy, Shanghai Jiao Tong University,
Shanghai 200240, China and Collaborative Innovation Center for Advanced
Microstructures, Nanjing 210093, China}

\affiliation{Department of Physics, Renmin University of China, Beijing 100872,
China }

\author{Rong Yu}

\affiliation{Department of Physics, Renmin University of China, Beijing 100872,
China }

\affiliation{Department of Physics and Astronomy, Shanghai Jiao Tong University,
Shanghai 200240, China and Collaborative Innovation Center for Advanced
Microstructures, Nanjing 210093, China}
\begin{abstract}
Motivated by recent experiments on the frustrated quantum magnetic
compound YbMgGaO$_{4}$, we study an effective spin model on triangular
lattice taking into account the effects of the spin-orbit coupling.
We determine the classical ground-state phase diagram of this model,
which includes a 120$^{\circ}$ N\'{e}el and two collinear antiferromagnetic
phases. In the vicinity of the phase boundary between the N\'{e}el and
collinear phases, we identify three intermediate non-collinear antiferromagnetic
phases. In each of them the magnetic moments are ordered at multiple
incommensurate wave vector $\mathbf{Q}$ values. We further study
the effects of quantum fluctuations in this model via a linear spin-wave
theory. We find that the spin excitation gap of the non-collinear multi-$\mathbf{Q}$ antiferromagnetic state
is finite but can be vanishingly small, and this state is 
unstable to a spin liquid phase under strong quantum fluctuations in some large
$|J_{z\pm}|$ regime.
\end{abstract}
\maketitle

\section{Introduction}

Frustrated magnets can hold exotic states of matter, such as a quantum
spin liquid (QSL) in which the spin rotational and time reversal symmetries
are preserved down to the temperature of absolute zero.\cite{Balents}
In the search of QSL, the triangular antiferromagnet is one of the
most well studied frustrated systems. By disturbing the 120$^{\circ}$
long-range antiferromagnetic order 
of the Heisenberg model
with certain tuning parameters, various QSL states on triangular lattice
have been proposed.\cite{Misguich99,Motrunich05,Kaneko14,LiCampbell15,Watanabe14}
Alternatively, strong spin-orbit coupling (SOC) may introduce non-Heisenberg
exchange couplings and is found to be an effective way in stabilizing
some exotic quantum states, including a QSL, of frustrated magnets.\cite{Kitaev,compass,Kugel-Khomskii,DMKagome}
Recently, a new triangular antiferromagnet with strong SOC, YbMgGaO$_{4}$,
has been proposed to be a candidate compound of gapless QSL.\cite{LiChenZhang2015,LiChen2015}
In this material, it is shown that the strong SOC gives rise to large
spin and spatial entangled anisotropic interactions, which are suggested
to be crucial in stabilizing a QSL ground state.\cite{LiChenZhang2015,LiChen2015}

An effective model Hamiltonian for YbMgGaO$_{4}$ has been proposed
in Ref.~\cite{LiChenZhang2015}. It contains strong anisotropic non-Heisenberg
interactions due to SOC. But little is known for this model. Even
the classical phase diagram of this model has not been well studied.
And it is still unclear whether these anisotropic non-Heisenberg terms
in the model would provide sufficiently strong quantum fluctuations
to stabilize a QSL, and how would such a state be relevant to the
likely QSL phase observed in experiments. To address these questions,
we investigate the ground-state phase diagram and spin excitations
of this model. We determine the classical ground-state phase diagram
by 
numerical optimization and a modified Luttinger-Tisza (LT) method.
The phase diagram contains a $120^{\circ}$ N\'{e}el antiferromagnetic
(AFM) phase, two collinear AFM phases, and three novel incommensurate
non-collinear AFM phases. In these incommensurate phases, the magnetic
moments are ordered at multiple $\mathbf{Q}$ wave vectors. By using
the linear spin-wave theory, we find that all these classical magnetic
phases survive in the presence of weak quantum fluctuations. We further
calculate the spin-wave excitation in the non-collinear multi-$\mathbf{Q}$
phase and find the spin excitation gap of this state is finite but can be vanishingly small. When
the quantum fluctuations are strong, we find that a spin liquid phase
can be stabilized in the phase diagram.

The paper is organized as follows: In Sec.~\ref{Sec:Model}, we present
the general effective spin model and outline the methods we used to
study its ground state and spin excitations. In Sec.~\ref{Sec:GS},
we determine the classical ground-state phase diagram of this model
by using a numerical zero-temperature energy optimization with the
aid of a modified LT method, and show that non-collinear multi-$\mathbf{Q}$
phases are stabilized in certain regimes of the phase diagram. In
Sec.~\ref{Sec:SpinWave}, we show the spin excitations within the
linear spin-wave calculations and the correction of the quantum fluctuations
to the ground-state phase diagram. We further discuss the implication
of the model and our results to the YbMgGaO$_{4}$ in Sec.~\ref{Sec:Discussion}.
Finally we draw conclusions in Sec.~\ref{Sec:Conclusion}.




\section{Model and Methods}

\label{Sec:Model}


In YbMgGaO$_{4}$, because of the strong spin-orbit coupling (SOC),
the electrons of the Yb$^{3+}$ ion are in a state of total angular
momentum $J=7/2$. The crystal field then splits it into a series
of Kramers doublets. At low temperatures, only the lowest Kramers
doublet is relevant and the system can be described by a model of
interacting effective spin-1/2 magnetic moments. Due to the separation
between the two Yb layers by the nonmagnetic Mg/GaO$_{5}$ layers,
the interlayer superexchange coupling between the effective moments
are very weak. We then neglect this interlayer exchange coupling,
and define the model on a two-dimensional triangular lattice.

The Hamiltonian of this model reads~\cite{LiChenZhang2015}
\begin{eqnarray}
H & = & \sum_{\langle ij\rangle}[J_{zz}S_{i}^{z}S_{j}^{z}+J_{\pm}(S_{i}^{+}S_{j}^{-}+S_{i}^{-}S_{j}^{+})\nonumber \\
 & + & J_{\pm\pm}(\gamma_{ij}S_{i}^{+}S_{j}^{+}+\gamma_{ij}^{*}S_{i}^{-}S_{j}^{-})\label{eq:ham}\\
 & - & \frac{iJ_{z\pm}}{2}(\gamma_{ij}^{*}S_{i}^{+}S_{j}^{z}-\gamma_{ij}S_{i}^{-}S_{j}^{z}+\langle i\leftrightarrow j\rangle)]\nonumber
\end{eqnarray}
Here $\mathbf{S}_{i}$ refers to the effective spin-1/2 magnetic moment,
and $J_{zz}$, $J_{\pm}$, $J_{\pm\pm}$, and $J_{z\pm}$ are exchange
couplings between nearest neighbor moments. In this paper, we are
interested in the case $J_{zz}>0$, which is relevant to the YbGaMgO$_{4}$
compound~\cite{LiChenZhang2015}. The coefficients $\gamma_{ij}$
are defined on each bond of the triangular lattice which take the
value $1$, $e^{i\frac{2\pi}{3}}$ and $e^{-i\frac{2\pi}{3}}$ for
$\pm\mathbf{a}_{1}$, $\pm\mathbf{a}_{2}$ and $\pm\mathbf{a}_{3}$
nearest-neighor bond directions, respectively. See Fig. \ref{Fig:1}(a).
The SOC couples the rotational symmetry in the spin space to that
in the real-space. This lowers the symmetry of the model from SU(2)
to $D_{3d}$. Therefore, the model is non-Heisenberg, with spin and
spatial anisotropic exchange couplings described by $J$'s and $\gamma_{ij}$.
Due to the effect of SOC, this Hamiltonian has only \emph{discrete}
time-reversal and $D_{3d}$ point group symmetries, but the ground
state may still contain some emergent continuous symmetry, as will
be discussed in detail below.




A powerful way to investigate the classical groundstate configuration
of spin models is the Luttinger-Tisza method~\cite{LuttingerTisza46}.
In this approach, one first performs the Fourier transformation for
$\mathbf{S}_{j}$,
\begin{equation}
\mathbf{S}_{j}=\sqrt{\frac{1}{N}}\sum_{\mathbf{k}}\mathbf{S}_{\mathbf{k}}e^{i\mathbf{k}\cdot\mathbf{R}_{j}},
\end{equation}
where the sum is taken in the first Brillouin zone. The Hamiltonian
in Eq.~\eqref{eq:ham} can then be rewritten to a tensor form
\begin{equation}
H=\sum_{\mathbf{k}}\mathbf{S}_{\mathbf{k}}^{*}\cdot\mathsf{J}_{\mathbf{k}}\cdot\mathbf{S}_{\mathbf{k}},
\end{equation}
where $\mathsf{J}_{\mathbf{k}}$ is a real symmetric tensor, taking
into account the symmetry of the model, and $\mathbf{S}_{\mathbf{k}}^{*}$
refers to the complex conjugate of $\mathbf{S}_{\mathbf{k}}$. It
is then diagonalized to be
\begin{equation}
H=\sum_{\mathbf{k}\mu}\mathrm{\omega}_{\mathbf{k}\mu}S_{\mathbf{k}\mu}^{*}S_{\mathbf{k}\mu},\label{eq:HamDiag}
\end{equation}
where $S_{\mathbf{k}\mu}=\mathbf{S}_{\mathbf{k}}\cdot\hat{\mathbf{e}}_{\mathbf{k}\mu}$,
$\mathrm{\omega}_{\mathbf{k}\mu}$ and $\hat{\mathbf{e}}_{\mathbf{k}\mu}$
are corresponding eigenvalues and orthorgonal eigenvectors of the
tensor $\mathsf{J}_{\mathbf{k}}$. Meanwhile, the local constraint
of the constant spin magnitude at an arbitrary site $j$,
\begin{equation}
\mathbf{S}_{j}\cdot\mathbf{S}_{j}=S^{2},\label{eq:LT_cons0}
\end{equation}
yields the equivalent hard constraints on $\mathbf{S}_{\mathbf{k}}$
for any wave vector $\mathbf{q}$:
\begin{equation}
\frac{1}{N}\sum_{\mathbf{k}}\mathbf{S}_{\mathbf{k}}\cdot\mathbf{S}_{\mathbf{q}-\mathbf{k}}=S^{2}\delta_{\mathbf{qG}},\label{eq:LT_cons1}
\end{equation}
where $\mathbf{G}$ is a reciprocal lattice vector. Also, since $\mathbf{S}_{j}$
are real vectors, each Fourier component must satisfy the relation
\begin{equation}
\mathbf{S}_{\mathbf{k}}^{*}=\mathbf{S}_{-\mathbf{k}}\label{eq:real_cons}
\end{equation}
In the original LT method, one minimizes the energy in Eq.~\eqref{eq:HamDiag}
under a released global constraint
\begin{equation}
\frac{1}{N}\sum_{\mathbf{k}}|\mathbf{S}_{\mathbf{k}}|^{2}=S^{2},\label{eq:LT_cons1-1}
\end{equation}
\emph{i.e.}, by taking $\mathbf{q}=\mathbf{G}$ in Eq.~\eqref{eq:LT_cons1}.
If the corresponding spin configuration of the minimum turns out to
satisfy Eq.\eqref{eq:real_cons} and all local constraints in Eq.\eqref{eq:LT_cons1}
as well, it must be the true physical ground state.

This method works well for conventional Heisenberg or XXZ models in
some parameter regimes. However, it has been showed that the LT method
failed to produce the physical ground state of the Hamiltonian in
Eq.~\eqref{eq:ham} because those hard constraints in Eq.~\eqref{eq:LT_cons1}
cannot be all satisfied simultaneously~\cite{Chen}. The deep underlying
reason is that the tensor $\mathsf{J}_{\mathbf{k}}$ of the Hamiltonian
contains only very low discrete symmetries, which will be discussed
in Appendix A.

To obtain the classical ground state of this model, we perform numerical
zero-temperature energy minimization of spin configurations in large
clusters. We find that besides the ordinary 120$^{\circ}$ N\'{e}el and
collinear phases discovered in the previous work, in the vicinity
of the N\'{e}el-collinear phase boundary, there exists three new phases
in which spins are ordered at multiple incommensurate $\mathbf{Q}$
points. We denote these phases as ``multi-$\mathbf{Q}$'' phases.
These multi-$\mathbf{Q}$ phase properties and the subtle phase transition
to collinear phase can be well produced in a modified LT approach,
by taking into account all the constraints in Eq.~\eqref{eq:LT_cons1}.
More details of this method is given in Appendix A.

To study the spin excitations and the effects of quantum fluctuations
to the classical ground states, we apply a linear spin-wave theory\cite{Toth2015,Petit2011,Wallace1962}
in real space by performing a local rotation on each spin $\mathbf{S}_{i}$.
The dynamical structure factor are calculated using the spinW codecs\cite{Toth2015}.
Details of the spin-wave approach is given in Appendix B.

\section{Classical ground-state phase diagram and the multi-$\mathbf{Q}$
state}

\label{Sec:GS}


\subsection{The phase diagram}

The model in Eq.~\eqref{eq:ham} has a rich phase diagram even for
classical spins. Let us first take a look at a special case where
$J_{zz}=2J_{\pm}-2J_{\pm\pm}\equiv J_{H}$ and $J_{z\pm}=0$. In this
case, Eq.~\eqref{eq:ham} reduces to a Heisenberg-120$^{\circ}$-compass
model~\cite{compass}
\begin{equation}
H=\sum_{\langle ij\rangle}(J_{H}\mathbf{S}_{i}\cdot\mathbf{S}_{j}+J_{c}S_{i}^{a}S_{j}^{a}),\label{eq:ham_compass}
\end{equation}
where $J_{c}=4J_{\pm\pm}$ and $a$ refers to the direction of the
bond $\langle ij\rangle$. To simplify the discussion, let us define
$\alpha=J_{c}/(J_{H}+J_{c})$.

\begin{figure}[h!]
\includegraphics[scale=1.4]{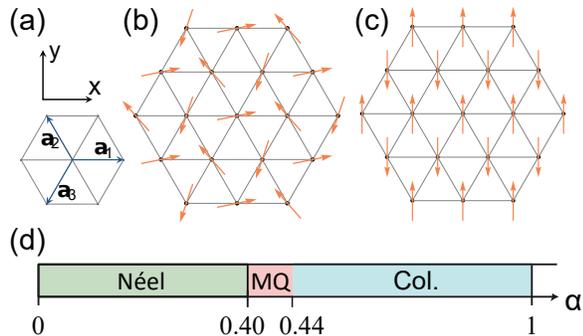} 

\caption{(a): Definition of the coordinate system and the nearest-neighbor
bonds. (b),(c): Spin patterns of the N\'{e}el and the collinear phases.
(d): The classical ground-state phase diagram of the Heisenberg-120$^{\circ}$-compass
model. Here the collinear and the multi-$\mathbf{Q}$ phases correspond
to the collinear II and multi-$\mathbf{Q}$ II phases in the generic
phase diagram of Fig.~\ref{Fig:2}(a) at $J_{z\pm}=0$, respectively.}

\label{Fig:1}
\end{figure}

It is known that in the Heisenberg limit ($\alpha=0$), the ground
state of this model is the 120$^{\circ}$ N\'{e}el AFM state\cite{Singh92,Capriotti99},
in which all spins lie in the plane of the lattice. While in the compass
limit ($\alpha=1$), the ground state is a collinear AFM state\cite{Wu2008},
in which all spins order ferromagnetically along one bond direction
but antiferromagnetically along the other two. See Fig. \ref{Fig:1}(b)(c).

Knowing the phases in the two limiting cases, we optimize the total
energy in large clusters to explore the ground state of a general
coupling $\alpha$. We find that the N\'{e}el state remains to be the classical
ground state for $\alpha<0.40$. Although the Hamiltonian has only discrete symmetry when the system is away from the Heisenberg point at $\alpha=0$, in the N\'{e}el state
the spin configurations still have degenerate energies under a global
rotation in the spin space with an arbitrary angle $\phi$ about the
$z$ axis. This is an example of an emergent $U(1)$ symmetry of the
ground state. As $\alpha$ further increases, we find an incommensurate
noncollinear AFM 
for $0.40<\alpha<0.44$, as shown in Fig.~\ref{Fig:1}(d). This state is denoted as the multi-$\mathbf{Q}$ state as the magnetic moments are ordered at multiple wave vectors in this state. Here we describe the phase diagram, and defer the discussion on the
nature of the multi-$\mathbf{Q}$ state to Sec.~\ref{Sec:MQ}. At
$\alpha\approx0.40$, we find a first-order transition between the
N\'{e}el AFM and the multi-$\mathbf{Q}$ state, while at $\alpha\approx0.44$,
the system undergoes a second-order transition from the multi-$\mathbf{Q}$ phase to the collinear AFM states.

\begin{figure}[h!]
\includegraphics[scale=0.3]{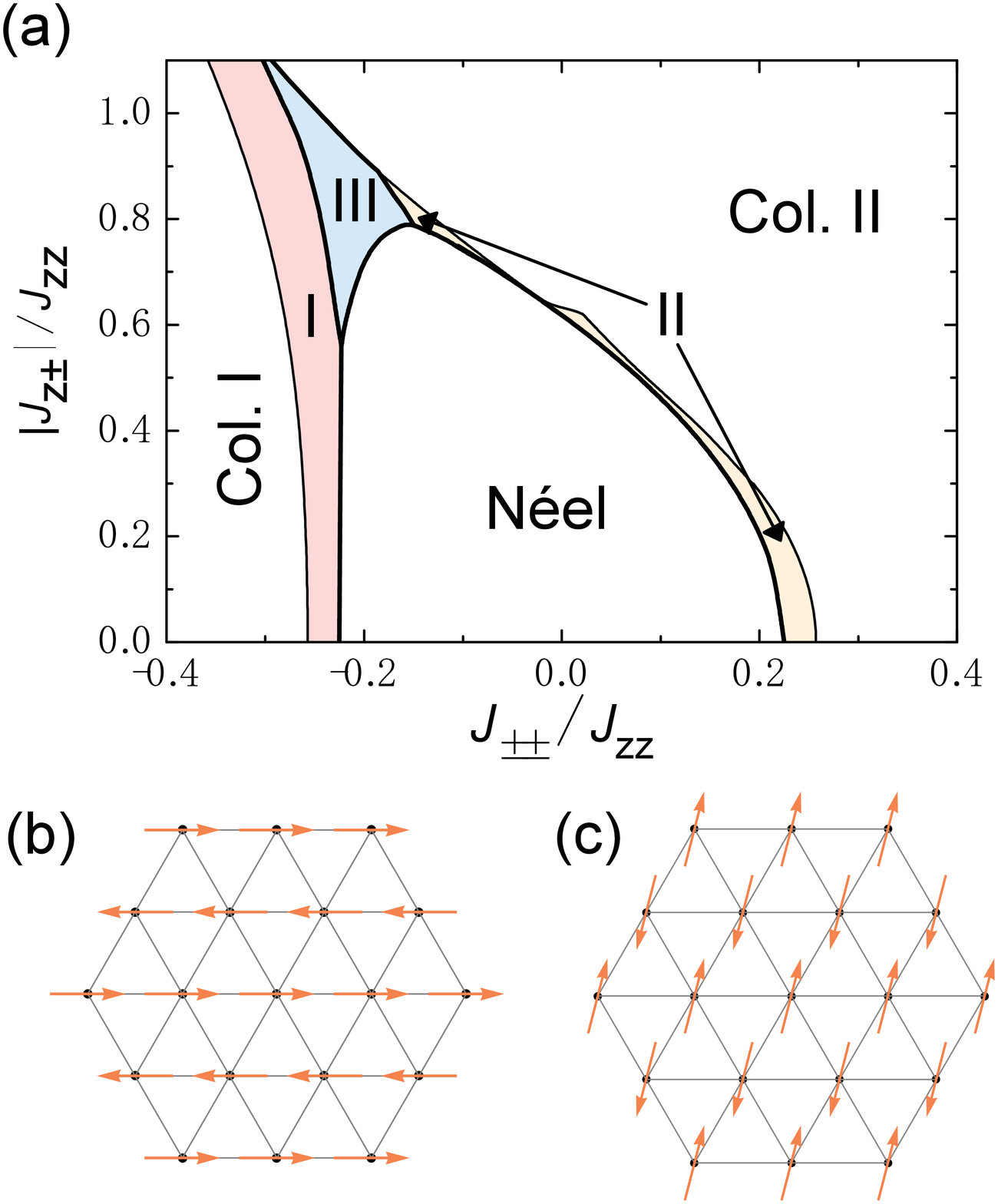} 

\caption{(a): Classical ground-state phase diagram of the generic spin-orbit
model defined in Eq.~\eqref{eq:ham}. Thicker and thinner curves
refer to first- and second-order transitions, respectively. Colored
regimes labeled as I, II, III correspond to the three multi-$\mathbf{Q}$
phases discussed in the text. 
Sketch of real-space spin patterns of the the two collinear states.
The N\'{e}el phase is as same as the one shown in Fig.~\ref{Fig:1}(b).}

\label{Fig:2}
\end{figure}

Compared to the Heisenberg-120$^{\circ}$-compass model, the full
model in Eq.~\eqref{eq:ham} contains additional anisotropic terms.
In our paper, the ratio $J_{\pm}/J_{zz}$ is fixed to be $0.9$, an
input from the experimental results of the YbGaMgO$_{4}$ single crystals~\cite{LiChenZhang2015}.
But the phase diagram is similar for other $J_{\pm}/J_{zz}>0.5$ values.
Our numerical energy optimization result reveals that the ground-state
phase diagram still contains N\'{e}el, collinear, and multi-$\mathbf{Q}$
phases. The emergent $U(1)$ symmetry of the N\'{e}el phase also exists
for this model. The multi-$\mathbf{Q}$ phase lies in between the
collinear and the N\'{e}el AFM phases, as shown in Fig.~\ref{Fig:2}(a).
When $J_{z\pm}=0$, the groundstates (so do the Hamiltonians) with
opposite signs of $J_{\pm\pm}$ are connected by $90^{\circ}$ rotation
in spin space about the $z$ axis.

For $J_{z\pm}\neq0$, the two collinear phases with opposite $J_{\pm\pm}$
values are no longer equivalent. In the collinear I phase spins are
still aligned along one bond direction while in the collinear II phase
spins are aquired to have finite $z$ components so as to further
minimize the energy. Two multi-$\mathbf{Q}$ states at either side
of the N\'{e}el phase (which we denote as multi-$\mathbf{Q}$ I and multi-$\mathbf{Q}$
II phases, respectively) are not equivalent either. Nor do they lie
in the $xy$ plane. But both of them 
coplanar. Also, we find that the multi-$\mathbf{Q}$ I to collinear
I and multi-$\mathbf{Q}$ II to collinear II transitions are second-order,
while all other transitions are first-order. N\'{e}el AFM state can be
stabilized at a vast range of $J_{\pm\pm}$ and $J_{z\pm}$ values.

When $J_{z\pm}$ is large, 
phase(donated as multi-$\mathbf{Q}$ III phase) is stabilized on the
upper side of the N\'{e}el regime, where the spins are non-coplanar, and
have relatively large 
from the spin directions in the collinear II order. The phase transitions
between the multi-$\mathbf{Q}$ III phase to others are first-order.

\subsection{Nature of the multi-$\mathbf{Q}$ phase}

\label{Sec:MQ}

One can easily check that the collinear state satisfy local constraints
in Eq.~\eqref{eq:LT_cons1}, and for sufficiently large $|J_{\pm\pm}|$,
the minimum of the eigenvalue of the tensor $\mathsf{J}_{\mathbf{k}}$
is located at the wave vector $\mathbf{Q}_{0}=(0,2\pi\text{/\ensuremath{\sqrt{3}}})$,
the ordering wave vector of the collinear state. According to the
LT method, the collinear state must be the exact ground state of the
model in this regime. However, when $|J_{\pm\pm}|$ is decreased towards
the boundary between the collinear and the N\'{e}el states, the minimum
of the eigenvalues of the tensor $\mathsf{J}_{\mathbf{k}}$ is away
from the wave vector $\mathbf{Q}_{0}$, while the energy minimum produced
by the LT method no longer satisfy all local constraints. Therefore,
in this $|J_{\pm\pm}|$ regime, the LT method fails to give the correct
ground state configuration of the system.

By taking the numerical energy minimization analysis, we find that
as $|J_{\pm\pm}|$ decreases so that the minimum of the eigenvalues
of $\mathsf{J}_{\mathbf{k}}$ deviates from $\mathbf{Q}_{0}$, the
ground state of the system does not immediately change. The collinear
state remains to be the ground state at this stage. However, as $|J_{\pm\pm}|$
further decreases, 
depending on the ratio of $|J_{z\pm}/J_{zz}|$, the system may enter the intermediate
multi-$\mathbf{Q}$ state via either a first- or a second-order transition,
as shown in Fig. \ref{Fig:2}(a). We find that these multi-$\mathbf{Q}$
phases can be well reconstructed by introducing \textit{finite} Fourier
components $\mathbf{S}_{\mathbf{Q}}$'s on multiple $\mathbf{Q}$'s
based on the original collinear states so as to minimize the energy
on the premise of satisfying local constraints (\ref{eq:LT_cons1-1}).
The detail of the process is given in the Appendix A.

\begin{figure*}[!h]
\includegraphics[scale=0.3]{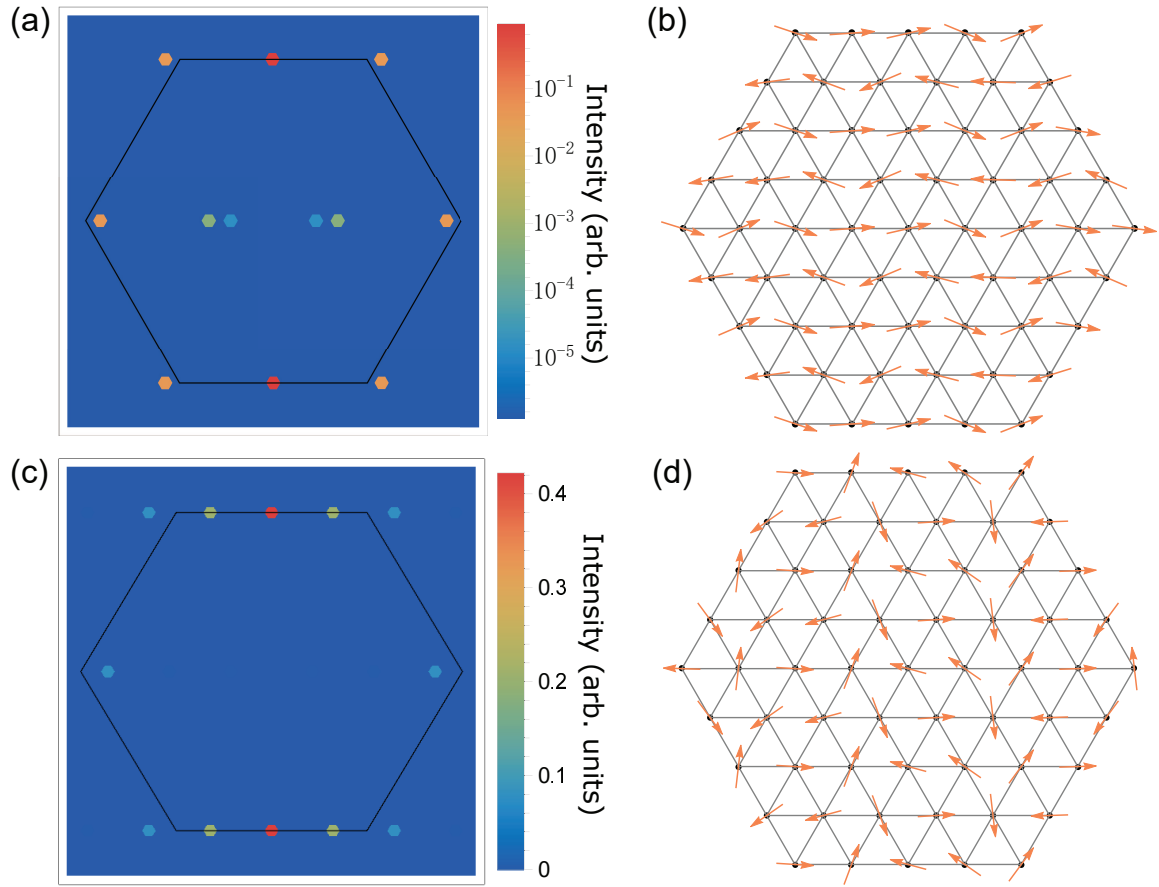} 

\caption{Color maps of the static structure factor and the sketchs of the real-space
spin pattern (projected to the $xoy$ plane) of: the coplanar multi-$\mathbf{Q}$
I state (in (a),(b)) and the non-coplanar multi-$\mathbf{Q}$ III
state (in (c),(d)). Here the model parameters we take for the multi-$\mathbf{Q}$
I state are $J_{\pm\pm}=-0.2165$, $J_{z\pm}=0$, and for the multi-$\mathbf{Q}$
III state are $J_{\pm\pm}=-0.19$, $J_{z\pm}=0.85$.}
\label{Fig:3}
\end{figure*}

Here we summarize the key results. We find that in multi-$\mathbf{Q}$
states the magnetic moments are ordered at multiple wave vectors,
as shown in Fig.~\ref{Fig:3}(a)(c). The spin structure factor shows
a primary peak at wave vector $\mathbf{Q}_{0}$, the ordering wave
vector of the collinear state. Two secondary peaks are present at
incommensurate wave vectors $\pm\mathbf{Q}_{1}$ along some high symmetry
line. For multi-$\mathbf{Q}$ I/II states, their spectral weights
of $\pm\mathbf{Q}_{1}$ are in general about one order of magnitude
smaller than the primary one. Other finite peaks of the structure
factor, for example, the peaks $\mathbf{Q}_{2}=[2\mathbf{Q}_{1}-\mathbf{Q}_{0}]$,
are also present, as shown in Fig.~\ref{Fig:3}(a)(c). Here the symbol
$[\mathbf{k}]$ represents the equivalent $\mathbf{k}$ point in the
first Brillouin zone. In fact, we reveal that in order to satisfy
all local constraints, in principle one need to introduce \textit{finite}
Fourier components for an \textit{infinite} series of wave vectors
$\mathbf{Q}_{n}$. But their spectral weights decays exponentially
with increasing $n$. For example, here the spectral weight of the
peak at $\mathbf{Q}_{2}$ is already about several orders of magnitudes
smaller than that of the primary peak. In practice, for multi-$\mathbf{Q}$
I and II phases, the peaks at $\mathbf{Q}_{n}$ for $n>2$ can hardly
be detected and have no physical significance. Therefore, as a good
approximation of the ground state, the series can be truncated at
$n=2$. For multi-$\mathbf{Q}$ III phase where the weight of $\mathbf{Q}_{2}$
and $\mathbf{Q}_{1}$ have been comparable to $\mathbf{Q}_{0}$, since
the spectra weight of $\mathbf{Q}_{n}$ for $n>2$ is still small,
our perturbative construction are still qualitatively valid to produce
the spin configurations.

\begin{figure}[h!]
\includegraphics[scale=0.33]{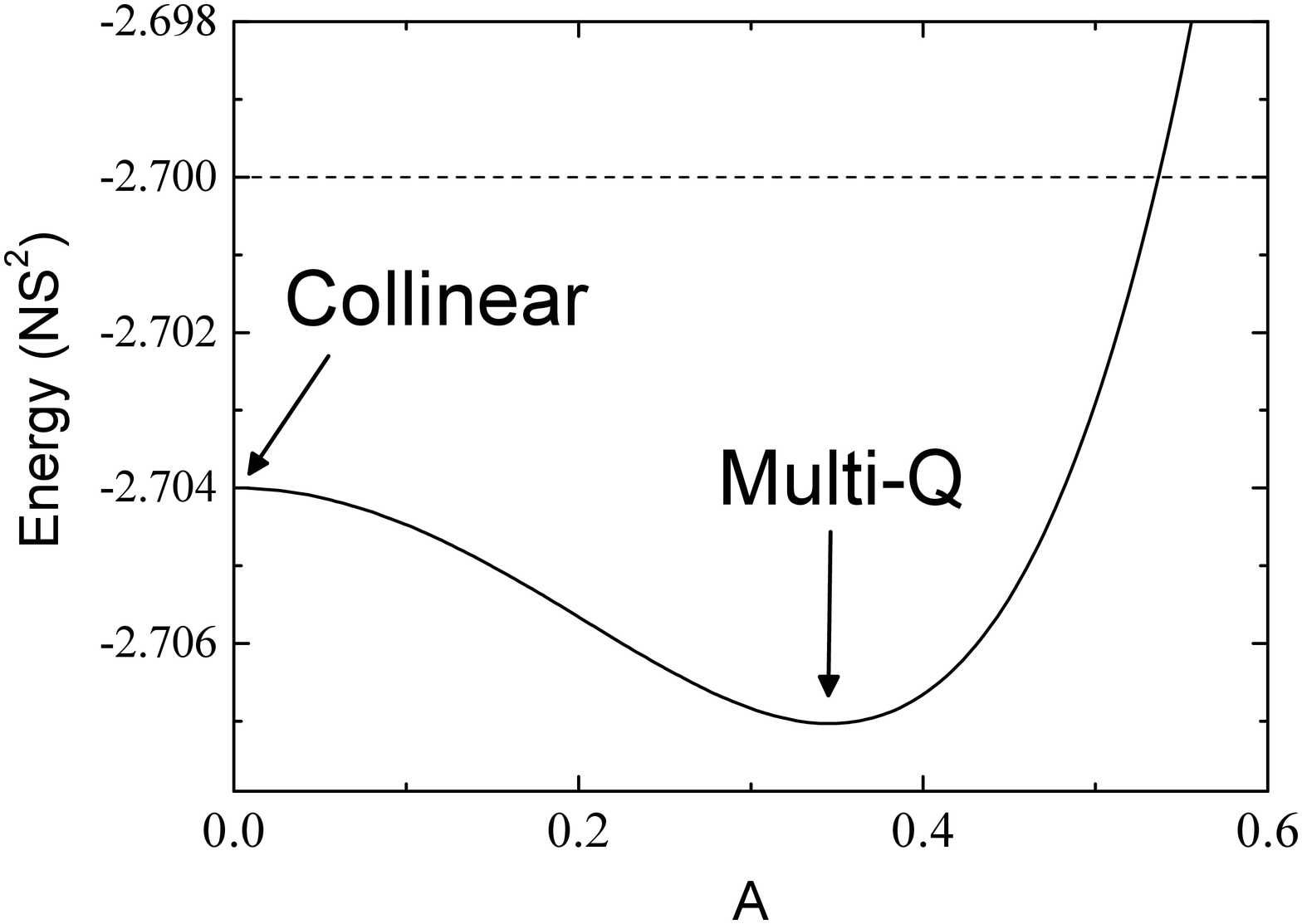} 

\caption{Energy versus modulation amplitude $A$ for the variational configuration
$\{\phi_{i}[A]\}$. Here we take the $J_{\pm\pm}=-0.226$ and $J_{z\pm}=0$.
Dashed line shows the energy of the N\'{e}el state.}

\label{Fig:4}
\end{figure}

Sketches of the real-space spin pattern of the multi-$\mathbf{Q}$
state are shown in Fig.~\ref{Fig:3}(b)(d). Take multi-$\mathbf{Q}$
I state for example: In a simple case $J_{z\pm}=0$, the spins all
lie in the plane of the lattice. The spin pattern exhibits additional
modulation on top of the collinear order, but does not form any spiral
order. By taking the above truncation, the angle $\phi_{i}$ that
a spin at site $\mathbf{R}_{i}$ deviates from the horizontal direction
can be expressed as $\phi_{i}=\sin^{-1}[A\sin(\mathbf{Q}_{1}\cdot\mathbf{R}_{i}+\phi_{0})]$
and $\phi_{i}=\pi-\sin^{-1}[A\sin(\mathbf{Q}_{1}\cdot\mathbf{R}_{i}+\phi_{0})]$
for alternating rows respectively, where $A=\frac{2|\mathbf{S}_{\mathbf{Q}_{1}}|}{N\sqrt{S}}$,
defined as the modulation amplitude, and $\phi_{0}$ is an arbitrary
angle related to the phase of the Fourier component $\mathbf{S}_{\mathbf{Q}_{1}}$.
Here $A$ scales the deviation to the collinear order. If we take
$A$ as a variational parameter and calculate the energy of the spin
pattern defined by $\phi_{i}[A]$ , we see (from Fig.~\ref{Fig:4})
that the energy of the collinear state (corresponding to $A=0$) is
a local maximum while the energy of the multi-$\mathbf{Q}$ state
(at $A\approx0.35$) is the minimum. This verifies that the multi-$\mathbf{Q}$
state, instead of the collinear one, is the ground state of the model
in the vicinity of the N\'{e}el-collinear phase boundary of the phase
diagram.

For multi-$\mathbf{Q}$ I and II states, despite the relatively large
deviation of multi-$\mathbf{Q}$ states from the collinear ones, we
can see from Fig.~\ref{Fig:4} that their energy difference are generally
neglectably small. While in large $|J_{z\pm}|$ regime, the non-coplanar
spin patten of the multi-$\mathbf{Q}$ III state can 
much more energy than 
state. Also, we can see that around the multi-$\mathbf{Q}$ energy
minimum, there exists large numbers of competing states with different
modulation amplitude $A$, wavevector $\mathbf{q}$ and phase $\phi_{0}$
close in similar energy scale. These competing states are 
in destablizing magnetic moments when thermal or quantum fluctuations
are switched on.

\section{Effects of quantum fluctuations and spin excitations in the multi-$\mathbf{Q}$
phase}

\label{Sec:SpinWave}

To investigate the spin excitations of the above antiferromagnetic
phases, we perform a linear spin wave(LSW) analysis. The detail of
the LSW method is given in the Appendix.

\begin{figure}[h!]
\includegraphics[scale=1.2]{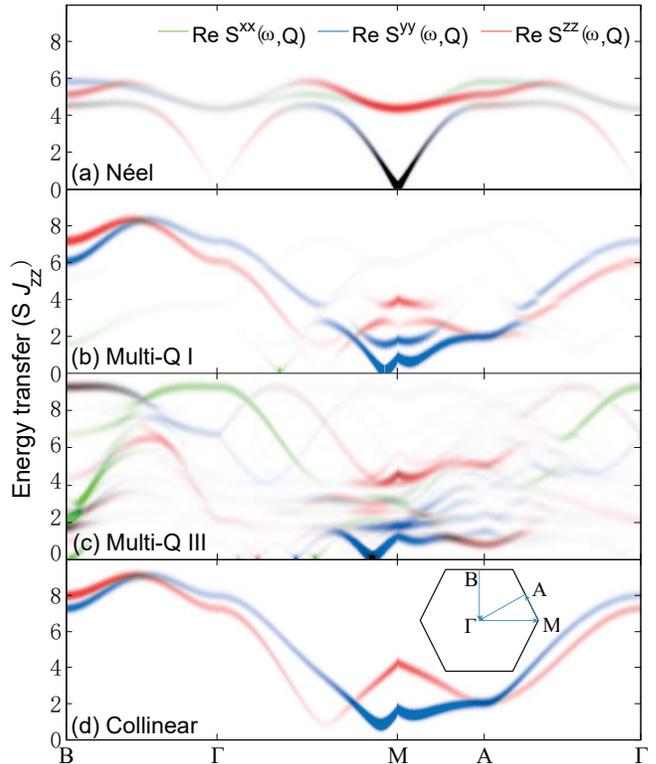} 

\caption{Dynamical structure factors of N\'{e}el, multi-$\mathbf{Q}$, and collinear
phases, respectively. The parameters for the N\'{e}el and multi-$\mathbf{Q}$
I phase are $J_{\pm\pm}=-0.2165$, $J_{z\pm}=0$, for the multi-$\mathbf{Q}$
III phase are $J_{\pm\pm}=-0.19$, $J_{z\pm}=0.85$, and for the collinear
phase are $J_{\pm\pm}=-0.28$, $J_{z\pm}=0$. The 
thickness of the color in each curve is proportional to the magnon spectral weight.}

\label{Fig:5}
\end{figure}

For the N\'{e}el, multi-$\mathbf{Q}$, and collinear states, the corresponding
dynamical structural factor, defined as,
\begin{equation}
S^{\mu\nu}(\mathbf{k},\omega)=\frac{1}{2\pi N}\sum_{ij}\int_{-\infty}^{+\infty}\mbox{d}t\,e^{i\mathbf{k}\cdot(\mathbf{r}_{i}-\mathbf{r}_{j})-i\omega t}\langle S_{i}^{\mu}S_{j}^{\nu}(t)\rangle,
\end{equation}
are shown in Fig.~\ref{Fig:5} for comparison. In the N\'{e}el state,
the spin excitation is gapless at M point of the Brilluion zone, as
a consequence of the emergent $U(1)$ symmetry mentioned in Sec.~\ref{Sec:GS}.
As for the collinear
state, the spin excitations are gapped, reflecting the discrete symmetry
of the model. The minimum of the spin-wave dispersion is located at
an incommensurate wave vector $\mathbf{Q}_{1}$ along some high symmetry
line. For moderate $J_{z\pm}$, when approaching to the collinear-to-multi-$\mathbf{Q}$
phase boundary by decreasing $|J_{\pm\pm}|$, the spin gap at $\mathbf{Q}_{1}$
drops to zero. Further decreasing $|J_{\pm\pm}|$, the spin-wave dispersion
of the collinear state near $\mathbf{Q}_{1}$ becomes imaginary, indicating
that the incommensurate magnon is condensed and the multi-$\mathbf{Q}$
phase develops. This is consistent with the collinear-to-multi-$\mathbf{Q}$
transition in the classical phase diagram. We also claim that along
the collinear-II-to-multi-$\mathbf{Q}$-II boundary in Fig.~\ref{Fig:2}(a),
collinear II phase is destablized at different $\mathbf{Q}_{1}$ points
for different parameters, which makes the boundary zig-zag.

Spin-wave excitation spectra of the multi-$\mathbf{Q}$ I/II state
look similar to those of the collinear phase in a large portion of
the Brilluion zone. But due to its complicated real-space spin structure,
the spectra of the multi-$\mathbf{Q}$ state contain multiple shadow
branches, which are most significantly seen near the M point. For
multi-$\mathbf{Q}$ III states, the spectra seems further scattered
due to the large modulation of incommensurate components. Particularly
in certain intermediate energy regime, sharp spin-wave dispersion
may not be well observed due to the various shaddow bands of magnons
that are associated with the complicated real-space spin pattern of
the multi-$\mathbf{Q}$ phase. Surprisingly, we find the spin gap
of multi-$\mathbf{Q}$ I/II and large portion of multi-$\mathbf{Q}$
III states is vanishingly small. This suggests existence of an (approximate) emergent
$U(1)$ symmetry. While this is not as obvious as in the N\'{e}el
phase, we can understand it in an intuitive way. Taking the Fourier
component $\mathbf{S}_{\mathbf{Q}_{1}}$ as a variational parameter,
near the energy minimum (corresponding to the multi-$\mathbf{Q}$
ground state), the energy depends weakly on the phase of $\mathbf{S}_{\mathbf{Q}_{1}}$.
The excitations along the phase direction (transverse direction to
the amplitude excitations) are then almost gapless, and develop an
approximate Goldstone mode at $\mathbf{Q}_{1}$. Nevertheless in some multi-$\mathbf{Q}$
III regime 
(close to the N\'{e}el AFM phase), the spin
excitation gap can be sizable.

\begin{figure*}[!h]
\includegraphics[scale=0.35]{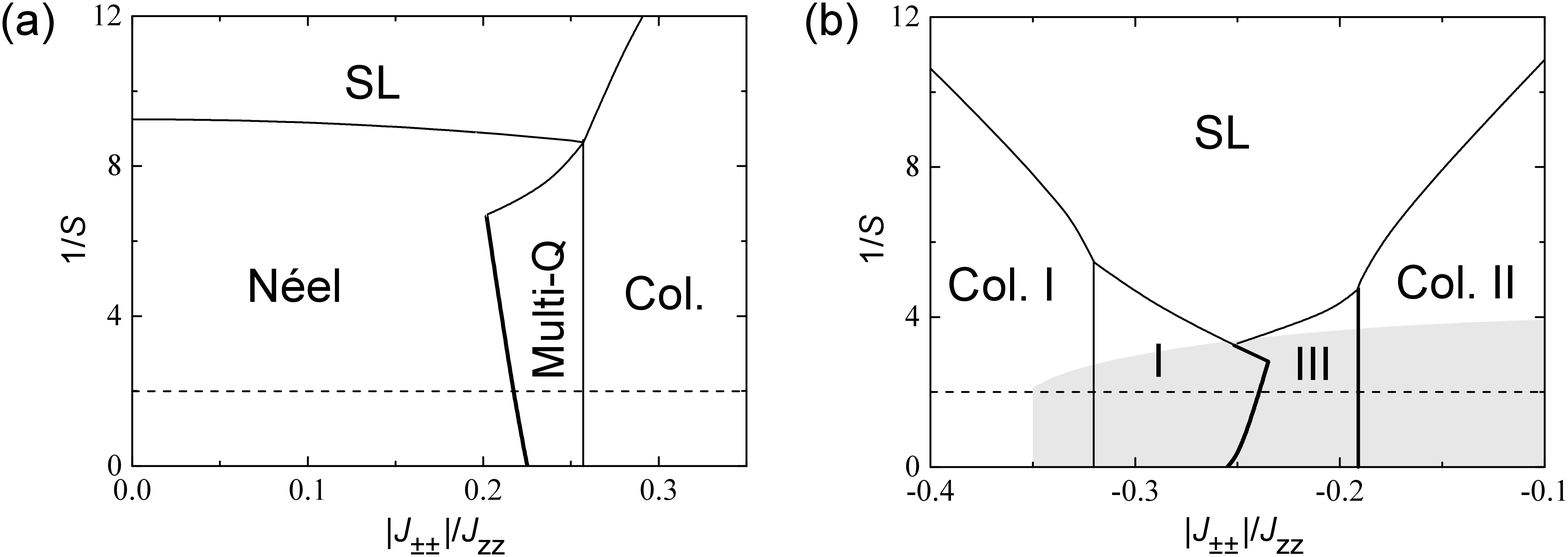} 

\caption{Phase diagrams taking into account the quantum correction from linear
spin-wave theory for (a): $J_{z\pm}=0$ and (b): $J_{z\pm}=0.9$.
Thinner solid curves correspond to second-order transitions and thicker
solid curves correspond to first-order transitions. The dashed line
marks $S=1/2$. In panel (b), the gray shading shows the regime where
the N\'{e}el state is a metastable state (which has an energy higher than
other ordered states).}
\label{Fig:6}
\end{figure*}

The LSW approach also allows us to exam the effects of quantum fluctuations
to the classical phases. Here we show the $1/S$ vs $J_{\pm\pm}/J_{zz}$
phase diagram in LSW theory at zero temperature in Fig. \ref{Fig:6}.
For $J_{z\pm}=0$ case, N\'{e}el, collinear, and multi-$\mathbf{Q}$ states
all survive weak to moderate quantum fluctuations. But the ordered
magnetic moments are reduced by quantum fluctuations. For N\'{e}el and
collinear states, the moment reduction is uniform for each sublattice,
while in multi-$\mathbf{Q}$ states, due to the complicated magnetic
structure, the moment reduction is inhomogeneous, and depends on the
neighboring environment of a spin in each sublattice. In each phase
the (largest) ordered moment reduction at is found to be $\lesssim0.16$,
so that the magnetic orders are robust even for $S=1/2$. In our calculation
a spin liquid phase can be stabilized for $1/S\gtrsim7$ where quantum
fluctuations are sufficiently strong.

When $|J_{z\pm}|$ is large, the phase diagram changes quite a bit,
as shown in Fig \ref{Fig:6}(b). We find that all phases become further
unstable against quantum fluctuations. Particularly, the ordered moment
reduction of the N\'{e}el state can be as large as $0.3$, and it can
only be stabilized as a metastable state since taking into account
the quantum corrections, its energy is higher than that of other ordered
states. 
moment reduction can be as large as around $0.3$. The multi-$\mathbf{Q}$
phase is completely unstable to a spin liquid at $1/S\approx3$. In
this sense, the system in this parameter regime is very close to a
spin liquid phase for $S=1/2$.

\section{Discussions}

\label{Sec:Discussion}

Magnetic order at multiple $\mathbf{Q}$ vectors usually exists in
systems with a complex lattice structure or competing exchange interactions
such that the magnetic unit cell contains more than one magnetic ion.~\cite{ThreeK}
The multi-$\mathbf{Q}$ phase we studied in this paper exists in simple
triangular lattice with nearest-neighbor exchange couplings. It is
induced by the anisotropic $J_{\pm\pm}$ interaction of the model,
which introduces strong competition between the 120$^{\circ}$ N\'{e}el
and the collinear phases. At low temperatures, the system attempts
to order at both wave vectors $\mathbf{Q}_{0}$ and $\mathbf{Q}_{{\rm N}}$,
and the multi-$\mathbf{Q}$ state is eventually stabilized as a compromise.
In fact, the existence of a large number of energetically competing
configurations around the multi-$\mathbf{Q}$ ground state is evidenced
by the shallow energy profile around the minimum in Fig.~\ref{Fig:3}.
The competition around the multi-$\mathbf{Q}$ ground state gives
rise to enhanced thermal fluctuations which can suppress the ordering
temperature of the multi-$\mathbf{Q}$ state. Actually, the reduction
of ordering temperature near the boundary between the N\'{e}el and collinear
phases has been observed in a recent Monte Carlo study~\cite{LiChen2015}.
But the multi-$\mathbf{Q}$ was not resolved in the Monte Carlo calculation
due to the limited system size and energy resolution.

Our LSW result shows that in some large $J_{z\pm}$ regime the system
can be close to QSL state for $S=1/2$ in this model. It should be
noted that in cases where $S$ is small and quantum fluctuation is
large, magnon interactions may significantly renormalize the system
and LSW approximation may become no longer valid. Therefore, it is
possible that magnon interactions may further supress the magnetic
order and drive the system towards a spin liquid. Also, as a semiclassical
approximation, spin wave theory considers quantum fluctuations above
only \textit{one} classically ordered state. In fact, quantum fluctuations
also allow tunnelling among different classical configurations with
similar energies. For multi-$\mathbf{Q}$ states, there exists large
numbers of competing states with similar energies (N\'{e}el, and other
multi-$\mathbf{Q}$ configurations with different $\mathbf{K}$ and
$\mathbf{S_{K}}$). Quantum(and thermal) tunneling among these states
may significantly destabilize the magnetic order. On the other hand,
the classical configuration of multi-$\mathbf{Q}$ state in real space
looks much more ``disordered'' than other conventional magnetic
phases, such as the N\'{e}el and collinear AFM states. Such a disordered
nature also shows up at the linear spin-wave level: the ordered moment
reduction is inhomogeneous. This makes the multi-$\mathbf{Q}$ state
susceptible to quantum fluctuations: once the quantum fluctuation
$1/S$ increases to the value such that the ordered moments of some
sites drop down to zero, the multi-$\mathbf{Q}$ state is distorted.
But the corresponding quantum disordered state can not be described
within the framework of a LSW approach. So it is possible that other
type of strong quantum fluctuations drive the system to a QSL via
distabilizing the multi-$\mathbf{Q}$ phase. In other words, the phase
diagram of the system under strong quantum fluctuations remains to
be explored, and it would be interesting to know the form of the elementary
excitations of the corresponding phase there.

For the YbMgGaO$_{4}$ compound, the seemingly divergent magnetic
susceptibility and the power-law behavior of the specific heat $C_{V}\sim T^{2/3}$
suggest absence of long-range magnetic order. Assuming that the system
can be described by the model Hamiltonian in Eq.~\eqref{eq:ham},
the superexchange couplings of the system have been recently estimated
from ESR measurements. It is found that $|J_{\pm\pm}/J_{zz}|\sim0.16$
and $|J_{z\pm}/J_{zz}|\sim0.04$.~\cite{LiChen2015} These parameters
suggest that the system is very close to the boundary between the
N\'{e}el and the multi-$\mathbf{Q}$ phase regime, as shown in Fig. \ref{Fig:1}(a).
But according to our LSW calculation, the ground state is still magnetically
ordered even for $S=1/2$. To reconcile the theory with the experimental
findings, on the one hand, other experimental measurements, such as
neutron and/or Raman scattering should be done to confirm or give
better estimates of the exchange couplings. It would be especially
important to accurately determine the value of the $J_{z\pm}$ coupling,
because our results show that a spin liquid state would be much easier
to be stabilized with a large $J_{z\pm}$ value. On the other hand,
other perturbations beyond the present model but likely existed in
the real materials, such as the longer ranged exchange couplings,
the ring exchange interaction, or disorder, may further disturb the
long-range magnetic order and drive the system toward a spin liquid~\cite{Misguich99,Motrunich05,Kaneko14,LiCampbell15,Watanabe14}.

In our model, the multi-$\mathbf{Q}$ states lie in large areas in the parameter
space of the phase diagram. One may be curious whether similar states exist in
other spin-orbit coupled system. Indeed, similar incommensurate
ordered states have been found in a number of theoretical models,
such as the Heisenberg-Kitaev model on triangular lattice, and Heisenberg
models on hyperhoneycomb and hyperkagome lattices.\cite{Daghofer2016,Becker2015,Lee2015,Lee2015a,Mizoguchi2016}
There are also some experimental evidences of these exotic magnetic
states\cite{Biffin2014,Biffin2014a}. However, to our knowledge, the
microscopic origin and physical properties of these states are not
yet well addressed. Given the similar magnetic structures of these incommensurate
states to the multi-$\mathbf{Q}$, they likely share the same
origin: as the symmetry is lowered by the SOC induced anisotropic interactions, the spin wave of the original commensurate magnetic ground state (denoted as the parent state) is distabilized, and the magnons condense at a nearby incommensurate wave vector. For example, in the Kitaev-Heisenberg
model on the triangular lattice, once a finite Kitaev exchange coupling is added
to the antiferromagnetic Heisenberg interaction, the N\'{e}el AFM ground state
immediately becomes unstable to an incommensurate $Z_{2}$ vortex crystal.
\cite{Becker2015,Daghofer2016}
This is clearly seen in the spin-wave spectrum of the N\'{e}el AFM state, which is destabilized around M point of the Brilluion zone as soon as the system goes away from the Heisenberg point. Interestingly, the nature of the incommensurate state is closely connected to the properties of its parent state. Still in the Heisenberg-Kitaev model, the parent state of the $Z_{2}$ vortex crystal state
is the three-sublattice 120$^{\circ}$ N\'{e}el AFM state, in which the order parameter space is SO(3), and $Z_2$ point topological defects are allowed.\cite{Kawamura1984} Therefore, the topologically nontrivial $Z_{2}$ vortice crystal is stabilized when its parent state is disturbed.\cite{Becker2015,Daghofer2016} However, for the model Hamiltonian in Eq.~\eqref{eq:ham}, the parent state of the multi-$\mathbf{Q}$ states are the two-sublattice collinear states. Therefore, the multi-$\mathbf{Q}$ states are topologically trivial. It would
be interesting to further explore whether such a scenario generally holds
for the magnetism in systems with strong spin-orbit coupling.

\section{Conclusions}

\label{Sec:Conclusion}

In summary, we investigate the semiclassical phase diagram of an effective
spin model describing the strongly spin-orbit coupled local moments
in YbMgGaO$_{4}$. We identify three novel incommensurate multi-$\mathbf{Q}$
antiferromagnetic states in the classical phase diagram of this model.
We study the spin excitations of these states using a linear spin-wave
theory, and find that the spin excitation spectra contain multiple
branches, and the excitation gap can be vanishingly small.
With the linear spin-wave theory, we further study the effects of quantum fluctuations on the classical magnetic orders, and find that all these phases are stable
under weak to moderate quantum fluctuations. A spin liquid phase is
stabilized for sufficiently strong quantum fluctuations when the anisotropic
exchange coupling $|J_{z\pm}|$ is large.

\section{Acknowledgement}

We would like to acknowledge useful discussions with G. Chen, P. Holdsworth,
Y.-D. Li, Y. S. Li, Z.-X. Liu, B. Normand, T. Roscilde, Q. M. Zhang,
Q. Luo and J. Zhao. This work was supported in part by the National Program on Key Research Project Grant number 2016YFA0300500 (X.Q.W. and R.Y.), by
the National Science Foundation of China Grant number 11574200 (X.Q.W.), by the National Science Foundation of China Grant number 11374361 and
the Fundamental Research Funds for the Central Universities and the
Research Funds of Remnin University of China Grant number 14XNLF08 (R.Y.).
R.Y. acknowledge the hospitality of the Physics Laboratory at ENS
de Lyon.

\appendix

\section{Construction of multi-$\mathbf{Q}$ states within a modified Luttinger-Tisza
approach}

States produced by the original LT method have single-$\mathbf{Q}$
structure, which means only $\mathbf{S}_{\pm\mathbf{Q}}$ is nonzero
among all Fourier components. One can easily check that, if the ordering
wave vector $\mathbf{Q}$ is time reversal invariant(TRI , which means
$\mathbf{Q}=-\mathbf{Q}+\mathbf{G}$) momentum, such state always
satisfy all local constraints \eqref{eq:LT_cons1} and produces physical
\textit{collinear} groundstates; if $\mathbf{Q}$ is non-TRI, by taking
$\mathbf{q}=2\mathbf{Q}$ in \eqref{eq:LT_cons1}, we have $\mathbf{S}_{\mathbf{Q}}\cdot\mathbf{S}_{\mathbf{Q}}=0$,
which implies that at least two component in $\mathbf{S}_{\mathbf{Q}}$
must be nonzero. Therefore, if the minimum of the eigenvalue of the
tensor $\mathsf{J}_{\mathbf{Q}}$ happens to be at least two-fold
degenerate, LT method can still produce physical \textit{helical}
groundstates, otherwise such method cannot produce physical groundstates
satisfying \eqref{eq:LT_cons1}.

In general, if a system has continuous $U(1)$ symmetry, in some parameter
regime the minimum of the eigenvalue of $\mathsf{J}_{\mathbf{Q}}$
have degeneracy, so LT method still works. However, the model we study
has only discrete symmetries, LT method immediately fails once the
minimum of the eigenvalue of $\mathsf{J}_{\mathbf{Q}}$ deviates $\mathbf{Q}_{0}$.
Nevertheless, such multi-$\mathbf{Q}$ groundstates can still be well
reconstructed in a modified version of LT method by introducing finite
Fourier components $\mathbf{S}_{\mathbf{Q}}$ on multiple $\mathbf{Q}$'s
based on the collinear states so as to minimize the energy on the
premise of satisfying all local constraints \eqref{eq:LT_cons1-1}.

For pure collinear state there is only one nonzero component $\mathbf{S_{Q_{0}}}$
where $\mathbf{Q}_{0}$ is the collinear ordering wave vector. If
there exists some (non-TRI) $\mathbf{Q}_{1}$ point where the eigenstate
of $\mathsf{J}_{\mathbf{\mathbf{Q}_{1}}}$ is lower than the minimum
eigenvalue of $\mathsf{J}_{\mathbf{\mathbf{Q}_{0}}}$, the system
may tend to partially condense at $\pm\mathbf{Q}_{1}$ points in order
to gain more energy. By taking \textbf{$\mathbf{q}=2\mathbf{\mathbf{Q}_{1}}$}
in \eqref{eq:LT_cons1} we find that we further need to introduce
$\mathbf{Q}_{2}=[2\mathbf{Q}_{1}-\mathbf{Q}_{0}]$ component in order
to satisfy local constraint, i.e. $\mathbf{S}_{\mathbf{Q}_{1}}\cdot\mathbf{S}_{\mathbf{Q}_{1}}+\mathbf{S}_{\mathbf{Q}_{0}}\cdot\mathbf{S}_{\mathbf{Q}_{2}}+\mathbf{S}_{\mathbf{Q}_{2}}\cdot\mathbf{S}_{\mathbf{Q}_{0}}=0$.

From the above equation we can see that the magnitude of $\mathbf{S}_{\mathbf{Q}_{2}}$
is about order of $|\mathbf{S}_{\mathbf{Q}_{1}}|^{2}/|\mathbf{S}_{\mathbf{Q}_{0}}|$.
In general, the eigenvalues of $\mathsf{J}_{\mathbf{\mathbf{Q}_{2}}}$
are much larger than the minimum ones of $\mathsf{J}_{\mathbf{\mathbf{Q}_{0}}}$
and $\mathsf{J}_{\mathbf{\mathbf{Q}_{1}}}$, so $|\mathbf{S}_{\mathbf{Q}_{2}}|$
is in principle very small in order not to cause too much energy penalty.
Following the same procedure, by taking \textbf{$\mathbf{q}=2\mathbf{\mathbf{Q}_{2}}$}
in \eqref{eq:LT_cons1} we find that other finite Fourier components,
say $\mathbf{S}_{\pm\mathbf{Q}_{3}}$, in order to satisfy such constraint.
In principle, following this way of construction, an infinite series
of finite $\mathbf{S}_{\mathbf{Q}_{n}}$ must be introduced in order
to satisfy \eqref{eq:LT_cons1} rigorously. However, their magnitude
decays exponentially and for $n>2$, these components are generally
too small to be detected (no greater than the order of $|\mathbf{S}_{\mathbf{Q}_{1}}|^{3}/|\mathbf{S}_{\mathbf{Q}_{0}}|^{2}$)
and have no physical significance, so we truncate the series at $n=2$.
Although the truncated configuration $\{\mathbf{S}_{\mathbf{Q}_{0}},\mathbf{S}_{\pm\mathbf{Q}_{1}},\mathbf{S}_{\pm\mathbf{Q}_{2}}\}$
do not exactly satisfy \eqref{eq:LT_cons1}, such an approximation
turns out to be very well reconstructing multi-$\mathbf{Q}$ configurations.

The groundstate is therefore obtained by minimizing

\begin{align}
E & \{\mathbf{S}_{\mathbf{Q}_{0}},\mathbf{S}_{\mathbf{Q}_{1}},\mathbf{S}_{\mathbf{Q}_{2}}\}=\mathbf{S}_{\mathbf{Q}_{0}}^{*}\cdot\mathsf{J}_{\mathbf{Q}_{0}}\cdot\mathbf{S}_{\mathbf{Q}_{0}}\nonumber \\
 & +2\mathbf{S}_{\mathbf{Q}_{1}}^{*}\cdot\mathsf{J}_{\mathbf{Q}_{1}}\cdot\mathbf{S}_{\mathbf{Q}_{1}}+2\mathbf{S}_{\mathbf{Q}_{2}}^{*}\cdot\mathsf{J}_{\mathbf{Q}_{2}}\cdot\mathbf{S}_{\mathbf{Q}_{2}}
\end{align}

within constraints (by taking $\mathbf{q}=\mathbf{G}$, $\mathbf{Q}_{0}+\mathbf{Q}_{1}$,
$2\mathbf{Q}_{1}$, $\mathbf{Q}_{1}+\mathbf{Q}_{2}$ in \eqref{eq:LT_cons1}
respectively)

\begin{eqnarray}
|\mathbf{S}_{\mathbf{Q}_{0}}|^{2}+|\mathbf{S}_{\mathbf{Q}_{1}}|^{2}+|\mathbf{S}_{\mathbf{Q}_{2}}|^{2} & = & NS^{2}\\
\mathbf{S}_{\mathbf{Q}_{0}}\cdot\mathbf{S}_{\mathbf{Q}_{1}}+\mathbf{S}_{\mathbf{Q}_{1}}\cdot\mathbf{S}_{-\mathbf{Q}_{2}} & = & 0\label{eq:q0q1}\\
\mathbf{S}_{\mathbf{Q}_{1}}\cdot\mathbf{S}_{\mathbf{Q}_{1}}+2\mathbf{S}_{\mathbf{Q}_{0}}\cdot\mathbf{S}_{\mathbf{Q}_{2}} & = & 0\\
\mathbf{S}_{\mathbf{Q}_{1}}\cdot\mathbf{S}_{\mathbf{Q}_{2}} & = & 0\label{eq:q1q2}
\end{eqnarray}

We can see from \eqref{eq:q0q1} and \eqref{eq:q1q2} that $\mathbf{S}_{\mathbf{Q}_{1}}\perp\mathbf{S}_{\mathbf{Q}_{2}}$
and approximately $\mathbf{S}_{\mathbf{Q}_{1}}\perp\mathbf{S}_{\mathbf{Q}_{0}}$
(as the magnitude of $\mathbf{S}_{-\mathbf{Q}_{2}}$ is generally
much smaller than $\mathbf{S}_{\mathbf{Q}_{0}}$). By taking the energy
optimization, we find that the energy minimum of multi-$\mathbf{Q}$
I/II phases satisfy $\mathbf{S}_{\mathbf{Q}_{2}}\parallel\mathbf{S}_{\mathbf{Q}_{0}}$,
i.e. all $\mathbf{S}_{\mathbf{Q}}$'s are in the same plane. So multi-$\mathbf{Q}$
I/II states are coplanar, with all spins lying in the plane spanned
by $\mathbf{S}_{\mathbf{Q}_{0}}$ and $\mathbf{S}_{\mathbf{Q}_{1}}$.
However, such relation do not satisfy for multi-$\mathbf{Q}$ III
phases, which implies that multi-$\mathbf{Q}$ III states are not
coplanar.

Also, we find that the phase of $\mathbf{S}_{\mathbf{Q}_{1}}$, which
is relavent to the spin configuration, do not effect the energy within
our approximation.

\section{Linear spin wave method}

Here we present our linear spin wave method which applies to systems
with periodic structure in classical configuration. Suppose the configuration
can be devided into $M$ sublattices. This method apparently works
for N\'{e}el(3 sublattices) and collinear(2 sublattices) order. For multi-$\mathbf{Q}$
states one can still apply such method if we carefully choose the
model parameters and the cluster size such that all ordering wavevectors
well matches the reciprocal lattice of the cluster.

The method is performed as following\cite{Toth2015,Petit2011,Wallace1962}.
Suppose that ground state has classical configuration $\left\{ \mathbf{n}_{ns}\right\} $
where $\mathbf{n}_{ns}$ is the unit vector pointing direction of
the spin at the site $i$ labeled by magnetic unit cell index $n$
and sublattice index $s$. Since the spin direction only depends on
the sublattice index $s$, i.e. $\mathbf{n}_{ns}=\mathbf{n}_{s}$.
For each $\mathbf{n}_{s}$ one can always find a rotation operation
$R_{s}\in SO(3)$ that rotates $\hat{z}$ to $\mathbf{n}_{s}$ direction,
i.e., $\mathbf{n}_{s}=R_{s}\hat{z}$.

Introduce $\mathbf{S}_{ns}=R_{s}\tilde{\mathbf{S}}_{ns}$, so each
$\tilde{\mathbf{S}}_{ns}$ has classical configuration ferromagnetically
aligned along $\hat{z}$ direction. Then we perform H-P transformation
for $\tilde{\mathbf{S}}_{ns}$.

\begin{eqnarray}
\tilde{S}_{ns}^{z} & = & S-b_{ns}^{\dagger}b_{ns}\nonumber \\
\tilde{S}_{ns}^{+} & = & \sqrt{2S-b_{ns}^{\dagger}b_{ns}}b_{ns}\\
\tilde{S}_{ns}^{-} & = & b_{ns}^{\dagger}\sqrt{2S-b_{ns}^{\dagger}b_{ns}}\nonumber
\end{eqnarray}

At the LSW level, $\mathbf{S}_{ns}$ can be expressed as

\begin{equation}
\mathbf{S}_{ns}=\sqrt{\frac{S}{2}}(\mathbf{u}_{s}^{*}b_{ns}+\mathbf{u}_{s}b_{ns}^{\dagger})+\mathbf{v}_{s}(S-b_{ns}^{\dagger}b_{ns})\label{eq:Sns}
\end{equation}
where $u_{s}^{\mu}=R_{s}^{\mu x}+iR_{s}^{\mu y}$, and $v_{s}^{\mu}=R_{s}^{\mu z}$
for $\mu=x,y,z$ components.

Take eq. \eqref{eq:Sns} into the Hamiltonian \eqref{eq:ham}, after
Fourier transformation

\begin{equation}
b_{ns}=\sqrt{\frac{M}{N}}\sum_{\mathbf{k}\in MBZ}b_{\mathbf{k}s}e^{i\mathbf{R}_{ns}\cdot\mathbf{k}}
\end{equation}

The Hamiltonian can be rewritten in terms of boson bilinears at the
LSW level

\begin{equation}
H=E_{0}+\frac{1}{2}\sum_{\mathbf{k}\in MBZ}[\Psi(\mathbf{k}){}^{\dagger}h(\mathbf{k})\Psi(\mathbf{k})-\frac{1}{2}\mbox{tr}\,h(\mathbf{k})]
\end{equation}

where $E_{0}$ is the classical energy, $\Phi(\mathbf{k})=\left[b_{\mathbf{k}1},\cdots,b_{\mathbf{k}M},b_{\mathbf{-k}1}^{\dagger},\cdots,b_{\mathbf{-k}M}^{\dagger}\right]^{T}$,
$h(\mathbf{k})$ is a $2M\times2M$ Hermitian matrix.

$H$ can be diagonalized via Bogoliubov transformation$\Psi(\mathbf{k})=T_{\mathbf{k}}\Phi(\mathbf{k})$
where $\Phi(\mathbf{k})=\left[\beta_{\mathbf{k}1},\cdots,\beta_{\mathbf{k}M},\beta_{\mathbf{-k}1}^{\dagger},\cdots,\beta_{-\mathbf{k}M}^{\dagger}\right]^{T}$
and $T_{\mathbf{k}}\in SU(M,M)$ in order to ensure bosonic commutation
rules of Bogoliubov quasiparticles $\beta$'s.

The dianonalized Hamiltonian reads

\begin{eqnarray}
H & = & E_{0}+\frac{1}{2}\sum_{\mathbf{k}\in MBZ}[\Phi(\mathbf{k}){}^{\dagger}E(\mathbf{k})\Phi(\mathbf{k})-\frac{1}{2}\mbox{tr}\,h(\mathbf{k})]\nonumber \\
 & = & E_{0}+E_{r}+\sum_{\mathbf{k}\in MBZ}\omega_{\mathbf{k}s}\beta_{\mathbf{k}s}^{\dagger}\beta_{\mathbf{k}s}
\end{eqnarray}
where $E(\mathbf{k})=\mbox{diag}[\omega_{\mathbf{k}1},\cdots,\omega_{\mathbf{k}M},-\omega_{-\mathbf{k}1},\cdots,-\omega_{-\mathbf{k}M}]$
and $E_{r}=\frac{1}{4N}\sum_{\mathbf{k}\in MBZ}\mbox{tr}\,[E(\mathbf{k})-h(\mathbf{k})]$
is the zero point energy correction due to quantum fluctuation.

Following \cite{Toth2015}, at zero temperature, the ordered moment
reduction for the s'th sublattice $\Delta m_{s}$ reads

\begin{eqnarray}
\Delta m_{s} & = & \frac{M}{N}\langle\sum_{n}b_{ns}^{\dagger}b_{ns}\rangle\nonumber \\
 & = & \frac{M}{N}\sum_{\mathbf{k}\in MBZ}(T_{\mathbf{k}}T_{\mathbf{k}}^{\dagger})_{s+M,s+M}
\end{eqnarray}

and the dynamical structrual factor take the form

\[
S^{\mu\nu}(\mathbf{k},\omega)=\frac{S}{2N}\sum_{s=1}^{M}[T_{\mathbf{k}}^{\dagger}\mathbf{U^{\mu}}(\mathbf{U^{\nu})}^{\dagger}T_{\mathbf{k}}]_{s+M,s+M}\delta(\omega-\omega_{\mathbf{k}s})
\]

where $\mathbf{U}^{\mu}=[u_{1}^{\mu},\cdots,u_{M}^{\mu},(u_{1}^{\mu})^{*},\cdots,(u_{M}^{\mu})^{*}]^{T}$
are vectors in $2M$ dimension.

\bibliographystyle{apsrev4-1}
%
\end{document}